\documentclass[preprint,superscriptaddress,floatfix,pra,amssymb,amsmath]{revtex4}
\usepackage[dvips]{graphicx}
\usepackage{epsfig}
\usepackage{tabularx}
\usepackage{color}
\usepackage[pdfpagemode=UseNone,colorlinks=true,linkcolor=blue,citecolor=blue]{hyperref}

\newcommand{\bh}{\hat{b}}
\newcommand{\bd}{\hat{b}^\dagger}
\newcommand{\nm}{\bar{n}}
\newcommand{\nth}{\bar{n}_{\mathrm{th}}}

\newcommand{\Geff}{\Gamma_{\mathrm{eff}}}
\newcommand{\Gopt}{\Gamma_{\mathrm{opt}}}
\newcommand{\Gpar}{\Gamma_{\mathrm{par}}}
\newcommand{\Gm}{\Gamma_{\mathrm{m}}}
\newcommand{\Gplus}{\Gamma_{\mathrm{+}}}
\newcommand{\Gminus}{\Gamma_{\mathrm{-}}}
\newcommand{\Rplus}{R_{+}}
\newcommand{\Rminus}{R_{-}}
\newcommand{\Omegam}{\Omega_{\mathrm{m}}}
\newcommand{\dW}{\delta\Omega}
\newcommand{\Dpump}{\Delta}

\newcommand{\Dshift}{\Delta_{\mathrm{shift}}}
\newcommand{\epsc}{\epsilon_{\mathrm{c}}}

\newcommand{\Det}{\mathcal{D}}

\renewcommand\Re{\operatorname{Re}}
\renewcommand\Im{\operatorname{Im}}

\begin{document}

\title{Quantum motion of a squeezed mechanical oscillator attained via a optomechanical experiment}

\author{P. Vezio}
\affiliation{European Laboratory for Non-Linear Spectroscopy (LENS), Via Carrara 1, I-50019 Sesto Fiorentino (FI), Italy}

\author{A. Chowdhury} 
\affiliation{CNR-INO, L.go Enrico Fermi 6, I-50125 Firenze, Italy}

\author{M. Bonaldi}
\affiliation{Institute of Materials for Electronics and Magnetism, Nanoscience-Trento-FBK Division,
 38123 Povo, Trento, Italy}
\affiliation{Istituto Nazionale di Fisica Nucleare (INFN), Trento Institute for Fundamental Physics and Application, I-38123 Povo, Trento, Italy}

\author{A. Borrielli}
\affiliation{Institute of Materials for Electronics and Magnetism, Nanoscience-Trento-FBK Division,
 38123 Povo, Trento, Italy}
\affiliation{Istituto Nazionale di Fisica Nucleare (INFN), Trento Institute for Fundamental Physics and Application, I-38123 Povo, Trento, Italy}

\author{F. Marino}
\affiliation{INFN, Sezione di Firenze}
\affiliation{CNR-INO, L.go Enrico Fermi 6, I-50125 Firenze, Italy}

\author{B. Morana}
\affiliation{Institute of Materials for Electronics and Magnetism, Nanoscience-Trento-FBK Division,
 38123 Povo, Trento, Italy}
\affiliation{Dept. of Microelectronics and Computer Engineering /ECTM/DIMES, Delft University of Technology, Feldmanweg 17, 2628 CT  Delft, The Netherlands}

\author{G. A. Prodi}
\affiliation{Istituto Nazionale di Fisica Nucleare (INFN), Trento Institute for Fundamental Physics and Application, I-38123 Povo, Trento, Italy}
\affiliation{Dipartimento di Matematica, Universit\`a di Trento, I-38123 Povo, Trento, Italy}

\author{P.M. Sarro}
\affiliation{Dept. of Microelectronics and Computer Engineering /ECTM/DIMES, Delft University of Technology, Feldmanweg 17, 2628 CT  Delft, The Netherlands}

\author{E. Serra}
\affiliation{Istituto Nazionale di Fisica Nucleare (INFN), Trento Institute for Fundamental Physics and Application, I-38123 Povo, Trento, Italy}
\affiliation{Dept. of Microelectronics and Computer Engineering /ECTM/DIMES, Delft University of Technology, Feldmanweg 17, 2628 CT  Delft, The Netherlands}

\author{F. Marin}
\email[Electronic mail: ]{marin@fi.infn.it}
\affiliation{CNR-INO, L.go Enrico Fermi 6, I-50125 Firenze, Italy}
\affiliation{European Laboratory for Non-Linear Spectroscopy (LENS), Via Carrara 1, I-50019 Sesto Fiorentino (FI), Italy}
\affiliation{INFN, Sezione di Firenze}
\affiliation{Dipartimento di Fisica e Astronomia, Universit\`a di Firenze, Via Sansone 1, I-50019 Sesto Fiorentino (FI), Italy}

\date{\today}
\begin{abstract}
We experimentally investigate a mechanical squeezed state realized in a parametrically-modulated membrane resonator embedded in an optical cavity. We demonstrate that a quantum characteristic of the squeezed dynamics can be revealed and quantified even in a moderately warm oscillator, through the analysis of motional sidebands. We provide a theoretical framework for quantitatively interpreting the observations and present an extended comparison with the experiment. 
A notable result is that the spectral shape of each motional sideband provides a clear signature of a quantum mechanical squeezed state without the necessity of absolute calibrations, in particular in the regime where residual fluctuations in the squeezed quadrature are reduced below the zero-point level.
\end{abstract}
\maketitle

\section{Introduction}

Quantum mechanics has proven to be very effective in describing the behavior of physical systems at the atomic and subatomic level and large part of our technology is based on quantum processes occurring on microscopic scales. Since the inception of the theory, a long debate began on which typically quantum properties are conserved or modified  at the quantum-to-classical boundary and, eventually, observed in the macroscopic reality. While an extensive literature exists on non-classical properties of molecules, atoms and their constituents, experiments demonstrating genuine quantum properties of macroscopic degrees of freedom are comparatively very few. This situation has greatly changed in the last years thanks to the progresses in the realization and control of cavity optomechanical systems, in which the radiation pressure coupling between optical and mechanical modes allows to manipulate their state at the quantum level \cite{reviewOM}. 

One major result of the field is the cooling of macroscopic oscillators close to their quantum ground state \cite{meenehan,riedinger,peterson} and the observation of quantum signatures in their motion. One of such non-classical features is the so-called motional sidebands asymmetry. The mechanical interaction of the oscillator with a probe electromagnetic field measuring interferometrically its position, produces in the latter two motional sidebands at $\pm \Omegam$ around its main optical frequency \cite{safavi2012,purdy2015,under,pete2016,sudhir,Qiu2018,Qiu2019}, where $\Omegam$ is the mechanical eigenfrequency.
The asymmetry originates from the fact that a mechanical oscillator in the quantum ground state can only absorb energy. Since the blue sideband can be associated to an anti-Stokes scattering implying an energy transfer from the oscillator to the field (absorption of a phonon and frequency up-conversion of a photon) and vice versa for the red (Stokes) sideband, even at finite temperature the former process is less probable. The spectral areas below the blue and red sidebands are proportional to $\nth$ and $\nth+1$ respectively, where $\nth$ is the mean thermal occupation number. In cavity optomechanical systems, the interplay between radiation pressure force acting on the oscillator and the delay in the intracavity field buildup generates an additional mechanical damping, allowing the effective cooling of the mechanical motion to a lower mean phonon number $\nm$ \cite{reviewOM}. The ratio between Stokes and anti-Stokes sidebands can be written as $R_0 = (\nm+1)/\nm$ and a deviation from unity of $R_0$ becomes measurable for low enough $\nm$. Such an imbalance provides a direct measure of the displacement noise associated with zero-point quantum fluctuations \cite{Khalili2012,Weinstein2014,Borkje2016} and allows for a direct calibration of the mean phonon occupation number, i.e., of the absolute temperature of the oscillator \cite{purdy2017,sudhir2017,Chowdhury2019}. 

The preparation of strongly non-classical states in such systems and the identification of specific quantum indicators would represent a relevant step further, both from the scientific and technological point of view. Recent microwave experiments with cooled nano-oscillators have produced a mechanical squeezed state where the fluctuations in one quadrature are reduced below the zero-point level \cite{Wollman2015,Pirkkalainen2015,Lecocq2015,Lei2016}. Such a noise reduction can actually play an important role in several applications, for instance in optimizing the sensitivity of the oscillator used as quantum sensor. Nevertheless, the quadrature spectra do not carry any distinctive feature of the non-classical nature of the phenomenon, which can be only inferred (and actually demonstrated) "a-posteriori" through accurate calibration procedures. 

In a recent experiment we realized a squeezed state of a macroscopic mechanical oscillator embedded in an optical cavity \cite{ChowPRL2020}. The squeezing is generated via parametric modulation of the oscillator spring constant at twice its resonance frequency \cite{Rugar1991}. In this conditions classical and quantum fluctuations in phase with the parametric driving are amplified, while those $90$-degrees out of phase are suppressed. As a result, the oscillator motion in the phase space is squeezed in one quadrature and amplified in the orthogonal one. Such a scheme had already been implemented in cavity opto-mechanical experiments with thermal oscillators, where the modulation of the spring constant is achieved by modulating the light frequency or intensity \cite{Pontin_PRL2014,Sonar2018}. 
%The oscillator is indeed subject to a position-dependent radiation pressure force  that modifies the mechanical resonance frequency (optical spring): in our experiment such modulation is obtained by a suitable combination of optical fields. 

Remarkably, in the presence of parametric modulation the motional sidebands assume a peculiar shape, related to the modified system dynamics, unveiling the quantum component of the squeezed oscillator motion. Here we further investigate this phenomenon extending the experimental measurements of \cite{ChowPRL2020} and developing a theoretical framework able to explain the observations. 

The paper is organized as follows. In Sec. \ref{Sec_teo} we introduce the theoretical background with a discussion of key concepts and the derivation of relevant equations. In Sec. \ref{Sec:setup} we describe the experimental setup. In Sec. \ref{Sec:exp} we present the experimental results and compare them with the theoretical predictions. In particular, we report a complete characterization of the sideband asymmetry, as a function of the strength of the parametric drive, occupation number and detuning of the cooling beam. Our results demonstrate that 
some degree of quantum squeezing occurs and can be observed and quantified through the analysis of the motional sidebands, 
even for a moderately cooled mechanical oscillator, in which thermal noise is dominating over quantum fluctuations. We also show theoretically that, when the residual fluctuations in the squeezed quadrature are reduced below the zero-point level, the sideband spectra provide a clear indication of a quantum mechanical squeezed state, without the necessity of absolute calibrations. The conclusions and future perspectives are presented in Sec. \ref{Sec:concl}.
\section{Theoretical background}
\label{Sec_teo}

The linearized evolution equations for the intracavity field operator $\delta\hat{a}$ and the mechanical bosonic operator $\hat{b}$, in the frame rotating at frequency $\omega_{\mathrm{L}}$, are \cite{reviewOM}
\begin{equation}
\delta \dot{\hat{ a}}=\bigg(  i\Delta-\frac{\kappa}{2}\bigg)\delta\hat{a}+ig_0\alpha(\hat{b}+\hat{b}^\dag)+\sqrt{\kappa}\,\delta\hat{a}_{\mathrm{in}}
\label{dta}
\end{equation}
\begin{equation}
\dot{\hat{b}}=\left(-i\Omegam^0-\frac{\Gm}{2}\right)\hat{b}+ig_0(\alpha^* \delta\hat{a}+\alpha \delta \hat{a}^{\dag})+\sqrt{\Gm}\,\hat{b}_{\mathrm{th}}
\label{dtb}
\end{equation}
where $\Delta=\omega_{\mathrm{L}}-\omega_{\mathrm{c}}$ is the detuning with respect to the cavity resonance frequency $\omega_{\mathrm{c}}$, $\kappa$ and $\Gm$ are the optical and mechanical decay rates, $\Omegam^0$ is the mechanical resonance frequency, $g_0$ is the single-photon opto-mechanical coupling rate, and $\alpha$ is the intracavity mean field.
The input noise operators are characterized by the correlation functions
\begin{eqnarray}
\langle\hat{a}_{\mathrm{in}}(t)\hat{a}_{\mathrm{in}}^{\dag}(t')\rangle & = & \delta(t-t')
\label{noise1} \\
\langle\hat{a}_{\mathrm{in}}^{\dag}(t)\hat{a}_{\mathrm{in}}(t')\rangle & = & 0
\label{noise2}  \\
\langle\hat{b}_{\mathrm{th}}(t)\hat{b}_{\mathrm{th}}^{\dag}(t')\rangle & = & (\bar{n}_{\mathrm{th}}+1)\,\delta(t-t')
\label{noise3}    \\
\langle\hat{b}_{\mathrm{th}}^{\dag}(t)\hat{b}_{\mathrm{th}}(t')\rangle & = & \bar{n}_{\mathrm{th}}\,\delta(t-t')
\label{noise4}
\end{eqnarray}
where $\bar{n}_{\mathrm{th}}$ is the thermal occupation number. 

In our experiment the input field is composed of two tones, whose frequencies are shifted by $\pm \Omegam$ around their mean value $\omega_{\mathrm{L}}$. The shift frequency $\Omegam \simeq \Omegam^0$  is tuned to the effective mechanical resonance frequency, modified by the opto-mechanical interaction. It is determined experimentally by observing the oscillator spectrum, while in the framework of this theoretical model it will be defined later in a self-consistent way (Eq. \ref{eq_Omegam}). The mean value of the input field has the form
\begin{equation}
\alpha_{\mathrm{in}} = \alpha_-^{\mathrm{in}} \, e^{-i(\omega_{\mathrm{L}}-\Omegam)t} +  \alpha_+^{\mathrm{in}} \, e^{-i(\omega_{\mathrm{L}}+\Omegam)t} \, .
\end{equation}
The intracavity mean field, in the rotating frame, is $\alpha=\alpha_- e^{i\Omegam t}+\alpha_+ e^{-i\Omegam t}$, with amplitudes
\begin{equation}
\alpha_{\pm}=	\alpha_{\pm}^{\mathrm{in}}\,\frac{\sqrt{\kappa_{\mathrm{in}}}}{-i(\Delta\pm \Omegam)+\kappa/2}
\label{eq_alpha}
\end{equation}
where $\kappa_{\mathrm{in}}$ is the input coupling rate. 

In the Fourier space, equation (\ref{dta}) can be written as
\begin{equation}
\begin{split}
\delta \tilde{a}(\Omega)=\frac{1}{-i\Omega-i\Delta+\kappa/2}  \bigg\lbrace  ig_0\bigg[\alpha_{-} \left(\tilde{b}(\Omega+\Omegam)+\tilde{b}^{\dag}(\Omega+\Omegam)\right)
+\alpha_{+}\left(\tilde{b}(\Omega-\Omegam)+\tilde{b}^{\dag}(\Omega-\Omegam)\right)
\bigg] \\ +\sqrt{\kappa}\;\delta \tilde{a}_{\mathrm{in}}(\Omega) \bigg\rbrace 
\end{split}
\label{Fdta}
\end{equation}
where we use $\tilde{O}$ to indicate the Fourier transformed of the operator $\hat{O}$, and  $\tilde{O}^{\dag}$ for the Fourier transformed of $\hat{O}^{\dag}$, therefore $\left(\tilde{O}(\Omega)\right)^{\dag}=\tilde{O}^{\dag}(-\Omega)$. 

When Eq. (\ref{Fdta}) and its Hermitian conjugate are replaced into Eq. (\ref{dtb}) in the Fourier space, the mean field 
factors $\alpha$ and $\alpha^*$ in the optomechanical coupling shift the $\delta\tilde{a}$ and $\delta\tilde{a}^{\dag}$ 
operators by $\pm\Omegam$, giving terms proportional to $\tilde{b}(\Omega)$, $\tilde{b}(\Omega\pm 2\Omegam)$, 
$\tilde{b}^{\dag}(\Omega)$, $\tilde{b}^{\dag}(\Omega\pm2\Omegam)$. The equation (\ref{dtb}) can thus be written as
\begin{equation}
\begin{split}
\big(-i\Omega+i\Omegam^0+\Gm/2\big)\tilde{b}(\Omega)=\\
-g_0^2\left[\mathcal{C}_1\tilde{b}(\Omega)+\mathcal{C}_2\tilde{b}(\Omega-2\Omegam)+\mathcal{C}_3\tilde{b}(\Omega+2\Omegam)
+\mathcal{C}_4\tilde{b}^{\dag}(\Omega)+\mathcal{C}_5\tilde{b}^{\dag}(\Omega-2\Omegam)+\mathcal{C}_6\tilde{b}^{\dag}(\Omega+2\Omegam)
\right]+\tilde{b}_{\mathrm{in}}(\Omega)
\end{split}
\label{Fdtb}
\end{equation}
where $\mathcal{C}_i$ are c-numbers and the source term is
\begin{equation}
\begin{split}
\tilde{b}_{\mathrm{in}}(\Omega)  = \sqrt{\Gm}\; \tilde{b}_{\mathrm{th}}(\Omega)+ig_0\sqrt{\kappa}
\bigg[\alpha^{*}_{-}\frac{\delta \tilde{a}_{\mathrm{in}}(\Omega-\Omegam)}{-i\Omega-i\Delta+i\Omegam+\kappa/2}
+\alpha^{*}_{+}\frac{\delta \tilde{a}_{\mathrm{in}}(\Omega+\Omegam)}{-i\Omega-i\Delta-i\Omegam+\kappa/2} +\\
+\alpha_{-}\frac{\delta \tilde{a}^\dag_{\mathrm{in}}(\Omega+\Omegam)}{-i\Omega+i\Delta-i\Omegam+\kappa/2}
+\alpha_{+}\frac{\delta \tilde{a}^\dag_{\mathrm{in}}(\Omega-\Omegam)}{-i\Omega+i\Delta+i\Omegam+\kappa/2} 
\bigg] \, .
\end{split} 
\label{btildein}
\end{equation}
The total input noise source described by Eq. (\ref{btildein}) includes thermal noise and back-action noise, the latter given by the terms into square brackets. 

We now restrict our analysis to weak coupling, in which case the opto-mechanical damping rate and frequency shift of the mechanical oscillator (whose expressions will be given later, in Eqs. (\ref{eq_Gopt}) and (\ref{eq_Omegam})) are much smaller than its resonance frequency. Therefore, in the equation (\ref{Fdtb}) we just consider the quasi-resonant components in the opto-mechanical coupling term. From the left hand side of Eq. (\ref{Fdtb}), we see that $\tilde{b}$ is peaked around $\Omegam$, while $\tilde{b}^{\dag}$ is peaked around $-\Omegam$. Therefore, the relevant terms in the optomechanical coupling, on the right hand side of Eq. (\ref{Fdtb}), are those proportional to $\tilde{b}(\Omega)$ and $\tilde{b}^{\dag}(\Omega-2\Omegam)$. Writing the explicit form of the $\mathcal{C}$ coefficients, the equation (\ref{Fdtb}) becomes
%\begin{center} 
\begin{equation}
\begin{split}
\big(-i\Omega+i\Omegam^0+\Gm/2\big)\tilde{b}(\Omega)\simeq\\
-g_0^2\bigg[|\alpha_{-}|^2\;\tilde{b}(\Omega) 
\bigg(\frac{1}{-i\Omega-i\Delta+i\Omegam+\kappa/2}-\frac{1}{-i\Omega+i\Delta-i\Omegam+\kappa/2}\bigg)+\\
|\alpha_{+}|^2\;\tilde{b}(\Omega)\bigg( \frac{1}{-i\Omega-i\Delta-i\Omegam+\kappa/2}- \frac{1}{-i\Omega+i\Delta+i\Omegam+\kappa/2} \bigg)+\\
\alpha^{*}_{-}\alpha_{+}\;\tilde{b}^\dag(\Omega-2\Omegam)\bigg(\frac{1}{-i\Omega-i\Delta+i\Omegam+\kappa/2}-\frac{1}{-i\Omega+i\Delta+i\Omegam+\kappa/2}\bigg)
\bigg]+\tilde{b}_{\mathrm{in}}(\Omega)
\end{split}
\label{btilde}
\end{equation}
%\end{center}  
In the right hand side of Eq. (\ref{btilde}), we notice the usual opto-mechanical effects of the two laser tones (first two terms inside square brackets), plus their coherent common interaction, proportional to the fields product $\alpha^{*}_{-}\alpha_{+}$, that originates the parametric squeezing. It can be directly calculated that this parametric effect is null for $\Delta =0$, i.e., when the two tones are equally shifted with respect to the cavity resonance. 

The standard opto-mechanical interaction is parametrized by the total optical damping rate $\Gopt$, defined as \cite{reviewOM} 
%\begin{center} 
\begin{equation}
\begin{split}
\Gopt=2g^2_0 \Re \bigg[
|\alpha_{-}|^2\; 
\bigg(\frac{1}{-i\Omega-i\Delta+i\Omegam+\kappa/2}-\frac{1}{-i\Omega+i\Delta-i\Omegam+\kappa/2}\bigg)+\\
|\alpha_{+}|^2\;\bigg( \frac{1}{-i\Omega-i\Delta-i\Omegam+\kappa/2}- \frac{1}{-i\Omega+i\Delta+i\Omegam+\kappa/2} \bigg)
\bigg]  \, ,
\end{split}
\label{eq_Gopt}
\end{equation}
%\end{center}
and by a frequency shift that determines the effective resonance frequency $\Omegam$ according to the equation
%\begin{center} 
\begin{equation}
\begin{split}
\Omegam = \Omegam^0 + g^2_0\Im\bigg[
|\alpha_{-}|^2\; 
\bigg(\frac{1}{-i\Omega-i\Delta+i\Omegam+\kappa/2}-\frac{1}{-i\Omega+i\Delta-i\Omegam+\kappa/2}\bigg)+\\
|\alpha_{+}|^2\;\bigg( \frac{1}{-i\Omega-i\Delta-i\Omegam+\kappa/2}- \frac{1}{-i\Omega+i\Delta+i\Omegam+\kappa/2} \bigg)
\bigg]  \, .
\end{split}
\label{eq_Omegam}
\end{equation}
%\end{center}
For an easier comparison with the experimental data, it is useful to define the total opto-mechanical coupling strength $g^2 = g_0^2 \left(|\alpha_{-}|^2+|\alpha_{+}|^2\right)$ and the the ratio between intracavity powers $\epsc =  |\alpha_{-}|^2/(|\alpha_{-}|^2+|\alpha_{+}|^2)$.  Using the quasi-resonant frequency condition $\Omega \simeq \Omegam$, the total damping rate $\Geff = \Gm + \Gopt$ can be written as
\begin{equation}
\label{eqGeff}
\begin{split}
\Geff = \Gm + g^2 \kappa & \left(\frac{\epsc}{\Dpump^2+\kappa^2/4}-\frac{\epsc}{(\Dpump-2\Omegam)^2+\kappa^2/4} \right. \\
& \left. +\frac{1-\epsc}{(\Dpump+2\Omegam)^2+\kappa^2/4}-\frac{1-\epsc}{\Dpump^2+\kappa^2/4} \right)  \, .
\end{split}
\end{equation} 
With the same condition and notation, Eq. (\ref{btilde}) simplifies to  
\begin{equation}
\big(-i\Omega+i\Omegam+\Geff/2\big)\tilde{b}(\Omega)=\\
-\frac{\Gpar}{2} e^{i\phi}\;\tilde{b}^\dag(\Omega-2\Omegam)+\tilde{b}_{\mathrm{in}}(\Omega)
\label{btildeshort}
\end{equation} 
where
\begin{equation}
\Gamma_{\mathrm{par}} = \frac{4 g^2 \sqrt{\epsc (1-\epsc)} \, \Dpump}{\Dpump^2+\kappa^2/4}
\label{eq_Gpar}
\end{equation}
%\begin{equation}
%\Gpar = \frac{4 g_0^2 |\alpha_{+}||\alpha_{-}|\Delta}{\Delta^2+\kappa^2/4}
%\end{equation}
and $\phi = \pi/2+\arg[\alpha^{*}_{-}\alpha_{+}]$. 

Moving to the frame rotating at $\Omegam$ by means of the transformation 
\begin{equation}
\hat{b}_R  =  \hat{b} \, e^{i\Omegam t}   \qquad
\hat{b}_R^\dag  =  \hat{b}^\dag e^{-i\Omegam t}
\end{equation}
and, for Fourier transformed operators,
\begin{equation}
\tilde{b}_R(\Omega)  =  \tilde{b}\,(\Omega+\Omegam)   \qquad
\tilde{b}^\dag_R(\Omega)  =  \tilde{b}^\dag(\Omega-\Omegam)  
\end{equation}
and defining the frequency with respect to the mechanical resonance as $\dW = \Omega - \Omegam$, Eq. (\ref{btildeshort}) and its Hermitian conjugate can be written in the form of the system of coupled linear equations 
\begin{equation}
\left( \begin{matrix}
-i\dW+\frac{\Geff}{2} & \frac{\Gpar}{2} e^{i\phi} \\ \frac{\Gpar}{2} e^{-i\phi} &-i\dW+\frac{\Geff}{2} 
\end{matrix}\right)\left( \begin{matrix}
\tilde{b}_R \\\tilde{b}^\dag_R \end{matrix}\right)=\left( \begin{matrix}\tilde{b}_{\mathrm{in}} \\\tilde{b}^\dag_{\mathrm{in}} \end{matrix}\right)   \, .
\end{equation}
The determinant of the system matrix is
\begin{equation}
\Det =  \bigg(-i\dW+\frac{\Gamma_{+}}{2} \bigg) \bigg(-i\dW+\frac{\Gamma_{-}}{2} \bigg)
\end{equation}
where
\begin{equation}
\Gamma_{\pm}=\Geff\pm \Gpar  
\end{equation}
and the solutions of the system can be written as 
\begin{equation}
\tilde{b}_{R}=	\frac{1}{\Det}\bigg[\bigg(-i\dW+\frac{\Geff}{2} \bigg)	\tilde{b}_{\mathrm{in}}-\frac{\Gpar}{2} e^{i\phi}\;	\tilde{b}^\dag_{\mathrm{in}}
\bigg]
\label{bR}
\end{equation}
\begin{equation}
\tilde{b}^\dag_{R}=	\frac{1}{\Det}\bigg[\bigg(-i\dW+\frac{\Geff}{2} \bigg)	\tilde{b}^\dag_{\mathrm{in}}-\frac{\Gpar}{2} e^{-i\phi}\;	\tilde{b}_{\mathrm{in}}   
\bigg] \,  .
\label{bdagR}
\end{equation}
The correlation function for the input noise source of Eq. (\ref{btildein}) is obtained from Eqs. (\ref{noise1})-(\ref{noise4}) by considering that $\langle\hat{O}(t)\hat{O}^{\dag}(t')\rangle=c \, \delta(t-t')$ implies $\langle\tilde{O}(\Omega)\tilde{O}^{\dag}(\Omega')\rangle=2 \pi c \, \delta(\Omega+\Omega')$:
\begin{eqnarray}
\frac{1}{2\pi}\langle\tilde{b}_{\mathrm{in}}(-\Omega)\tilde{b}_{\mathrm{in}}^{\dag}(\Omega)\rangle & = & \Gm (\bar{n}_{\mathrm{th}}+1) + A^+   \\
\frac{1}{2\pi}\langle\tilde{b}_{\mathrm{in}}^{\dag}(-\Omega)\tilde{b}_{\mathrm{in}}(\Omega)\rangle & = & \Gm \,\bar{n}_{\mathrm{th}} + A^-   \\
\frac{1}{2\pi}\left\langle \tilde{b}_{\mathrm{in}}(-\Omega) \tilde{b}_{\mathrm{in}}(\Omega)\right\rangle=
\frac{1}{2\pi}\left\langle \tilde{b}^\dag_{\mathrm{in}}(-\Omega) \tilde{b}^\dag_{\mathrm{in}}(\Omega)\right\rangle^\ast
& = & -g_0^2\kappa\frac{\alpha^{*}_{-}\alpha_{+}}{\Delta^2+\kappa^2/4}
\end{eqnarray}
where the Stokes and anti-Stokes rates due to the two field tones are \cite{reviewOM}
\begin{eqnarray}
A^- & = &  g_0^2\kappa\bigg[\frac{|\alpha_{-}|^2}{\Delta^2+\kappa^2/4}+\frac{|\alpha_{+}|^2}{(\Delta+2\Omegam)^2+\kappa^2/4}
	   \bigg]   \\
A^+  & = &  g_0^2\kappa\bigg[\frac{|\alpha_{-}|^2}{(\Delta-2\Omegam)^2+\kappa^2/4}+\frac{|\alpha_{+}|^2}{\Delta^2+\kappa^2/4}
	   \bigg]
\end{eqnarray}	
and it can be verified that $\Gopt = A^- - A^+$. 

The spectra of the Stokes and anti-Stokes motional sidebands are finally calculated from Eqs. (\ref{bR})-(\ref{bdagR}) using the correlation functions given above, and are respectively
\begin{eqnarray}
S_{\bd_R \bd_R} & = & \frac{1}{2\pi}\langle\tilde{b}_{R}(-\dW)\tilde{b}_{R}^{\dag}(\dW)\rangle  = \frac{\Geff}{2}\left[\frac{1+\nm-s/2}{\dW^2+\Gminus^2/4}+\frac{1+\nm+s/2}{\dW^2+\Gplus^2/4}\right]    \label{eq_left}\\
S_{\bh_R \bh_R} & = & \frac{1}{2\pi}\langle\tilde{b}_{R}^{\dag}(-\dW)\tilde{b}_{R}(\dW)\rangle  = \frac{\Geff}{2}\left[\frac{\nm+s/2}{\dW^2+\Gminus^2/4}+\frac{\nm-s/2}{\dW^2+\Gplus^2/4}\right]   \label{eq_right}
\end{eqnarray}  	
where we have introduced the squeezing parameter
$s = \Gpar/\Geff$ and the oscillator effective phonon number in the absence of parametric effect is 
\begin{equation}
\bar{n} = \frac{\Gm \,\bar{n}_{\mathrm{th}} + \Gopt\, \bar{n}_{BA}}{\Geff}
\end{equation}
with $\bar{n}_{BA} = A^+/\Gopt$. 

The spectral shape of each motional sideband departs from a simple Lorentzian peak, and it is indeed composed by the sum of two Lorentzian curves with the same center, but different amplitudes and widths.
The ratios between the areas of the broad and narrow Lorentzian components in the two sidebands are 
\begin{eqnarray}
\Rplus = \frac{\nm +1+s/2}{\nm -s/2}    \label{ratios1}\\
\Rminus = \frac{\nm +1-s/2}{\nm +s/2}   \label{ratios2}
\end{eqnarray}
respectively for the broader ($\Rplus$) and narrower ($\Rminus$) components. In the absence of parametric gain ($s=0$) the two ratios coincide, and we recover the usual sideband asymmetry result for a thermal state $\Rplus = \Rminus = R_0$. 

A generic quadrature $X_{\theta}$ of the oscillator is defined as $X_{\theta} = (e^{i\theta}\, \bh_R + e^{-i\theta}\, \bd_R)/2$. The quadrature operator can be calculated in the Fourier space from Eqs. (\ref{bR})-(\ref{bdagR}), obtaining
\begin{equation}
\tilde{X}_{\theta} = \frac{1}{2\Det}\left[e^{i\theta} \,\tilde{b}_{\mathrm{in}} \left(-i\dW+\frac{\Geff}{2}-\frac{\Gpar}{2} e^{-i(2\theta+\phi)}\right)+e^{-i\theta} \,\tilde{b}^{\dag}_{\mathrm{in}} \left(-i\dW+\frac{\Geff}{2}-\frac{\Gpar}{2} e^{i(2\theta+\phi)}\right)\right]  \, .
\end{equation}
The shape of the spectrum of $X_{\theta}$ can be written as the sum of two Lorentzian functions, with width $\Gamma_{+}$ and $\Gamma_{-}$. Single Lorentzian shapes characterize the quadratures defined by $(2\theta+\phi)=0$ and $(2\theta+\phi)=\pi$. These quadratures are defined in the following as $Y \equiv X_{-\phi/2}$ and $X \equiv X_{-\phi/2+\pi/2}$. Their operators are 
\begin{equation}
Y  =  \frac{e^{-i\phi/2}\;\tilde{b}_{\mathrm{in}} +e^{i\phi/2}\;\tilde{b}_{\mathrm{in}}^\dag}{2\left(-i\dW+\frac{\Gamma_{+}}{2}\right)}    \qquad
X  =  \frac{i\left( e^{-i\phi/2}\;\tilde{b}_{\mathrm{in}} -e^{i\phi/2}\;\tilde{b}_{\mathrm{in}}^\dag \right)}{2\left(-i\dW+\frac{\Gamma_{-}}{2}\right)}
\end{equation}
and the associated spectra are  
\begin{equation}
S_{YY}   =   \frac{\Geff (2 \bar{n} + 1)}{4\left(\dW^2+\frac{\Gplus^2}{4}\right)}   \qquad
S_{XX}   =   \frac{\Geff (2 \bar{n} + 1)}{4\left(\dW^2+\frac{\Gminus^2}{4}\right)}   \, .
\end{equation}

The integrals of the spectra give different variances in the two orthogonal quadratures $\sigma_Y^2 = \sigma_0^2/(1+s)$ and $\sigma_X^2 = \sigma_0^2/(1-s)$ with $\sigma_0^2 = (2 \bar{n} + 1)/4$. The oscillator is said to be in a squeezed state. 

To summarize, in a classical description the motion of the oscillator is described by commuting variables, and the spectra corresponding to the two motional sidebands must be identical. On the other hand, in a quantum-mechanical description, even if the oscillator is dominated by thermal noise (i.e. $\nm  \gg  1$), the sideband asymmetry is always present ($R_0>1$), being originated by the non-commutativity between its ladder operators. Of course, the effect is actually measurable only for moderately low occupation numbers $\nm$. 

In the presence of parametric modulation, when the system is in a squeezed state, the sideband ratios $\Rplus$ and $\Rminus$ differ not only from unity, but also from the ratio $R_0$ measured in a thermal state. Namely, the ratio is higher for the broadened Lorentzian component, while for the narrowed component it approaches unity  as $s \to 1$ (i.e., close to the parametric instability threshold).

Therefore, the quantum features of the oscillator motion can be brought out even for a state having a variance exceeding that of the ground state in any quadrature and, besides thermal noise, even for states that are not of minimal uncertainty (i.e., with $\langle X^2 \rangle \langle Y^2 \rangle  > 1/16$) as those created by parametric squeezing. 

In the following, we describe an experimental study of this effect and provide evidence of the realization of a non-classical state of the macroscopic mechanical oscillator.

\section{Experimental Setup}
\label{Sec:setup}

\begin{figure}[h]
\includegraphics[width=0.95\columnwidth]{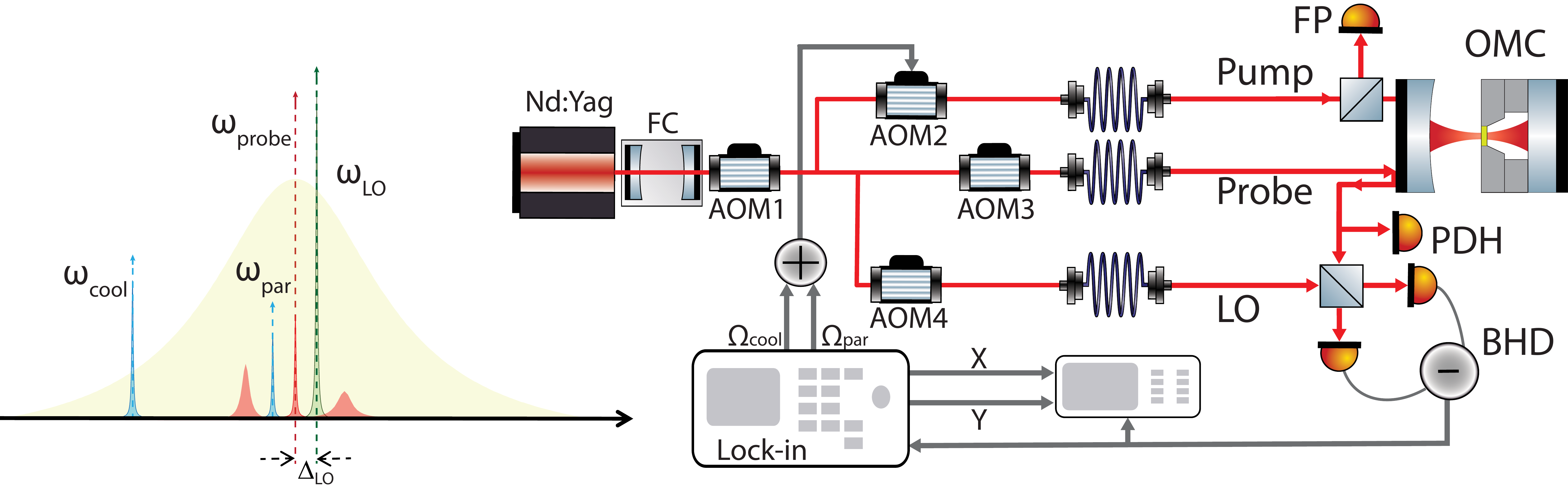}
\caption{Sketch of the experimental setup (see text) and conceptual scheme of the field frequencies. The LO is placed on the blue side of the probe ($\omega_{\mathrm{probe}}$) and detuned by $\Delta_{\mathrm{LO}}  \ll  \Omegam$, therefore the Stokes lines are on the red side of the LO, while the anti-Stokes lines are on the blue side. In the heterodyne spectra, they are located respectively at $\Omegam+\Delta_{\mathrm{LO}}$ (Stokes) and $\Omegam-\Delta_{\mathrm{LO}}$ (anti-Stokes).}
\label{fig1}
\end{figure}

A simplified scheme of the experimental setup and of the field frequencies used in the experiment is sketched in Figure \ref{fig1}. The mechanical oscillator is a circular SiN membrane with a thickness of $100\,$nm and a diameter of $1.64\,$mm, equipped with a specific on-chip structure that, working as a ``loss shield'' \cite{Borrielli2014,Borrielli2016,Serra2016,Serra2018}, reduces the coupling between the membrane and the frame and the consequent dissipation losses. Typical quality factors are of the order of a few millions for all the membrane modes, even at low frequencies, being limited by the intrinsic dissipation in SiN. 

In this work we exploit the (0,2) drum mode at $\Omegam/2\pi \simeq 530$ kHz, having a quality factor of $6.4 \times 10^6$ at cryogenic temperature. The membrane is placed inside a $3.92$~mm long cavity, $2$~mm far from the flat back mirror, thereby forming a ``membrane-in-the-middle'' setup \cite{Jayich2008}. The input mirror is concave with a radius of $50$~mm, which generates a waist of $70 \,\mu$m. 
%The membrane is kept roughly orthogonally to the incoming input beam. The orientation is important as slight deviation from the orthogonality can degrade the Finesse of the optical cavity by several factors. 
The cavity linewidth is $k/2\pi=1.9$~MHz corresponding to a finesse around $20000$. The vacuum optomechanical coupling factor is $g_{0}/2\pi=30$~Hz, determined by the overlap between the membrane mechanical mode and the beam waist. The optomechanical cavity is cooled down to $\sim 7\,$K in an helium flux cryostat. 

The light of a Nd:YAG laser is filtered by a Fabry-Perot cavity (FC) with a linewidth of $66$~kHz, frequency tuned by a first acousto-optic modulator (AOM1) and then split into three different beams.
A weak probe ($\sim 20 \,\mu$W), phase modulated at $13.3$~MHz, is kept resonant with the optomechanical cavity (OMC) using the Pound-Drever-Hall (PDH) technique with a servo loop exploiting AOM1 to correct fast fluctuations and a piezo-electric transducer to compensate slow changes of the cavity length. 

About 2 $\mu$W of the reflected probe are used for the PDH locking scheme, while most of the power ($\sim 10 \,\mu$W) is combined with a Local Oscillator (LO) beam ($\sim 2$~mW) and sent to a balanced detection (BHD). The LO frequency $\omega_{\mathrm{LO}}$ is blue-shifted with respect to the probe (namely, by $\Delta_{\mathrm{LO}}/2\pi = 11\,$kHz) by AOM4 to realize a low-frequency heterodyne detection \cite{Pontin2018a}. The BHD signal is acquired and off-line processed to study the motional sidebands and also sent to a lock-in amplifier and demodulated at frequency $\Omegam$. The two quadrature outputs of the lock-in are simultaneously acquired and off-line processed. 

The pump beam, orthogonally polarized with respect to the probe field, consists of two tones: the main one at frequency $\omega_{\mathrm{cool}}$ red detuned from the cavity resonance allows to cool down the mechanical motion. The second much weaker tone at a frequency $\omega_{\mathrm{par}}$, blue shifted with respect to the cooling beam by $\omega_{\mathrm{par}} - \omega_{\mathrm{cool}} = 2\Omegam$, realizes the parametric modulation of the oscillator generating the mechanical squeezing. The two tones are obtained by driving the AOM2 on the pump beam with the sum of two radiofrequency signals.
Reference spectra in the absence of resonant parametric effects are obtained by further shifting the modulation tone by 
$\Dshift \sim 2 \pi \times 12$ kHz. This shift is much larger than the mechanical width and much smaller than the cavity linewidth. This choice on one hand makes the coherent effect of the two tones negligible, and on the other hand keeps the cooling effect of the modulation tone almost constant.

During the experiment, the frequency difference between $\omega_{\mathrm{cool}}$ and $\omega_{\mathrm{par}}$ is periodically changed between $2\Omegam$ and $(2\Omegam+\Dshift)$ every $5$~s. 10~s long time series of the BHD signal are then acquired and the $5$~s segments are separated. Spectra with a resolution of 0.2~Hz are calculated from both segments, and averaged over 10 consecutive time series for the subsequent analysis. In this way we can accurately compare the spectra with and without parametric modulation even in the presence of slow variations in the system parameters.
In order to avoid the effects of possible long-term drifts during the measurements, all the radiofrequency signals used to drive the AOMs are phase-locked.

\section{Experimental Results}
\label{Sec:exp}

\begin{figure}
\begin{centering}
\includegraphics[width=0.8\columnwidth]{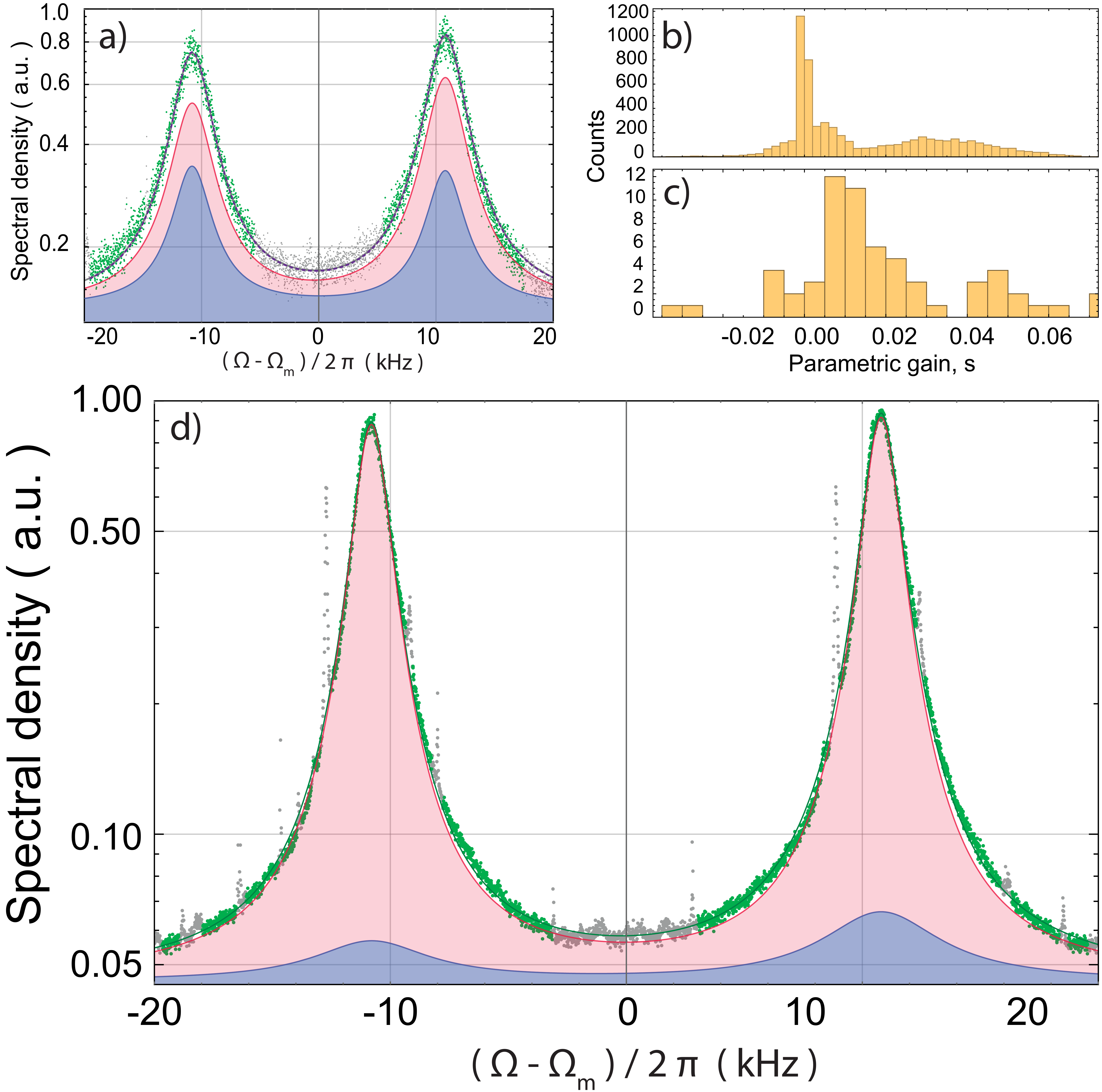}
\caption{(a, d) Heterodyne spectra around the ($0,2$) drum resonance at $\Omegam/2\pi \simeq 530\,$kHz showing the two motional sidebands separated by $\Delta_{\mathrm{LO}}/2\pi = 11\,$kHz. Each experimental spectrum (green symbols) is calculated from 10, $5\,$s long segments of the acquired time series. Gray symbols are used for data points excluded from the fitted regions. In (a) there is no resonant parametric drive, and the spectrum is fitted with one couple of Lorentzian curves (gray solid line) with equal width $\Geff$ and different amplitudes. For the same data we also show the fit curve with 
two couples of Lorentzian curves (violet dashed line) according to the expressions (\ref{eq_left}-\ref{eq_right}). The shaded (pink and light blue) regions shows the two Lorentzian contributions. Panel (c) shows in this case the statistical distribution for the parametric gain $s$, on 60 independent measurements. Panel (b) shows the statistical distribution of $s$ obtained with the same procedure on 6000 artificial, numerically generated spectra. In (d) (with resonant parametric drive) the parametric gain obtained from the fit with expressions (\ref{eq_left}-\ref{eq_right}) (here shown with a green solid line) is $s=0.53$.}
\label{figure2}
\end{centering} 
\end{figure}

As discussed in the previous sections, the analysis of the motional sidebands allows to explore the quantum components of the mechanical squeezed dynamics. Typical heterodyne spectra, displaying the two motional sidebands centered at frequencies $\Omegam \pm\Delta_{\mathrm{LO}}$, are shown in Fig. \ref{figure2}. Without any parametric modulation (panel (a)) the spectrum consists of a couple of Lorentzian curves, having the same width $\Geff$, but different areas. After the correction for the effect of the residual probe detuning \cite{Chowdhury2019}, their ratio $R_0$ provides a direct measurement of the mean occupation number $\nm$ through $R_0 = 1+1/\nm$. From this sideband asymmetry we infer $\nm = 5.8$ for the shown spectrum. 

In the presence of parametric modulation, the oscillator quadratures are no longer identical and each sideband is composed of two Lorentzian functions centered at the same frequency but with different linewidth and amplitude. The corresponding spectra for a given parametric gain $s$ are shown in Fig. \ref{figure2}d. The spectral peaks are fitted using Eqs. (\ref{eq_left}-\ref{eq_right}) where, in the two widths $\Gplus= \Geff (1+s)$ and $\Gminus=\Geff (1-s)$, $\Geff$ is fixed to the value extracted from the corresponding spectra in the absence parametric drive, while the parametric gain $s$ is left as free fitting parameter. We obtain a parametric gain of $s=0.53\pm 0.01$, where the quoted error corresponds to the standard deviation in 5 consecutive independent measurements, each one lasting 100~s. The pink and light blue areas plotted in Fig. \ref{figure2}d show the contributions to the motional sidebands from the squeezed $Y$-quadrature and amplified $X$-quadrature. We remark that the shape of each sideband is no longer represented by one single Lorentzian, and carries a distinctive signature of the squeezed oscillator motion.

In order to verify that the fitting procedure on the heterodyne spectra, assuming two couples of Lorentzian functions, is not biased, we have applied it to a set of $60$ independent spectra acquired in the absence of parametric drive, for different values of pump power and detuning. An example of the fitted Lorentzian components is shown in Fig. \ref{figure2}a (shaded pink and light blue regions). The statistical distribution of $s$, displayed in Fig. \ref{figure2}c, is characterized by a mean value of $0.038$ and a standard deviation of $0.046$, a result which is compatible with $s=0$ as expected. The standard deviation on $s$ is similar to those obtained in the presence of parametric drive. For a further check, we have generated artificial spectra with $s=0$ and the same parameters and signal-to-noise ratio of the experiment, and repeated the analysis. The statistical distribution of $s$ derived from the fits is shown in Fig. \ref{figure2}b, and displays a mean value of 0.014 and standard deviation of 0.019, figures similar to the experimental ones. It can be noticed that both the experimental and the artificial distributions are slightly asymmetric, with comparable skewness (respectively 1.12 and 0.7). This feature seems therefore related to the fitting procedure. We have not further studied this issue, but we infer that the analysis is reliable at the few hundredths level on $s$.  

\begin{figure}
\begin{centering}
\includegraphics[width=0.8\columnwidth]{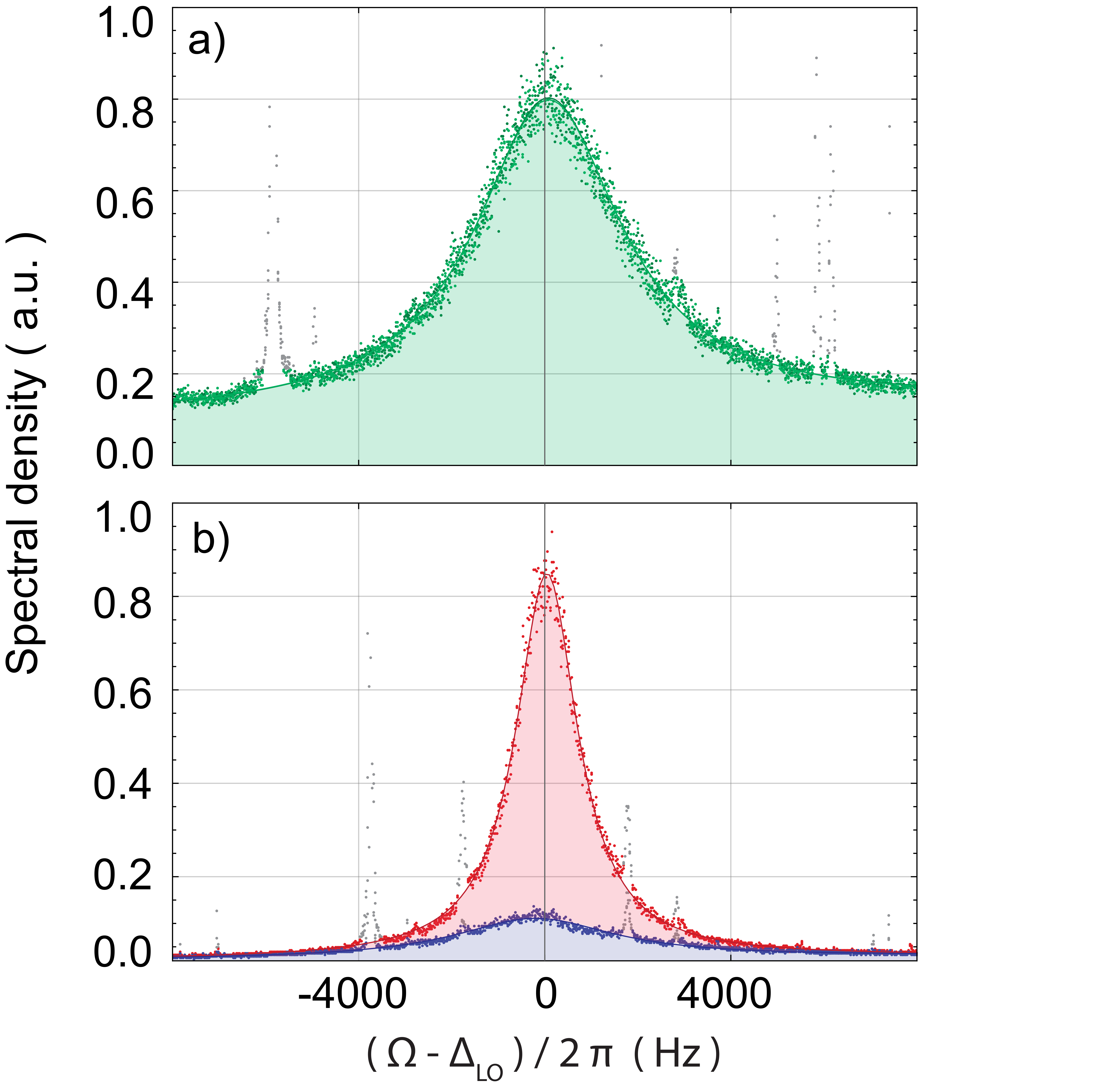}
\caption{Spectra of two orthogonal quadratures at the output of the lock-in amplifier. The demodulation phase is chosen to obtain respectively maximal ($X$) and minimal ($Y$) variance in the two signals, in the presence of resonant parametric modulation. a) The parametric tone is detuned from the resonant condition. The experimental signals of the two quadratures (light and dark green symbols) are not distinguishable, and one single Lorentzian fit is shown with a solid line. b) With resonant parametric drive, the spectra in the two quadratures (respectively red and blue symbols) are fitted with different single Lorentzian
curves (red and blue solid lines). gray symbols show spurious electronic peaks that are excluded from the fits.}
\label{figure_quadratures}
\end{centering} 
\end{figure}

The most straightforward method to show squeezing is the direct measurement of the variance in different quadratures, that are usually chosen by tuning the local oscillator phase in a homodyne detection. On the other hand, in a standard heterodyne setup the rapidly rotating phase difference between signal and local oscillator prevents the access to selected quadratures. This drawback can be overcome by controlling such phase difference \cite{Pontin2018a}. In our setup, all the oscillators are indeed phase locked, including the time base of a lock-in amplifier that demodulated the heterodyne signal at $\Omegam$. The spectrum of the lock-in output signal is a quadrature spectrum, centered at $\Delta_{\mathrm{LO}}$ and symmetrized, i.e., $\frac{1}{2} \left(S_{X_{\theta}X_{\theta}}(\omega-\Delta_{\mathrm{LO}})+S_{X_{\theta}X_{\theta}}(\omega+\Delta_{\mathrm{LO}})\right)$. The phase of the demodulator allows to tune $\theta$, and thus select the $Y$ and $X$ quadratures. Examples of the recorded spectra are shown in Fig. \ref{figure_quadratures}. The quadratures spectra are acquired and analyzed independently from the heterodyne signals, and the the analyses agrees for both the Lorentzian widths and the squeezing factor \cite{ChowPRL2020}.       

In the following we describe an extended characterization of the parametric squeezing as a function of different meaningful parameters, and compare the experimental results with the model reported in Section \ref{Sec_teo}. 

\begin{figure}
\begin{centering}
\includegraphics[width=0.8\columnwidth]{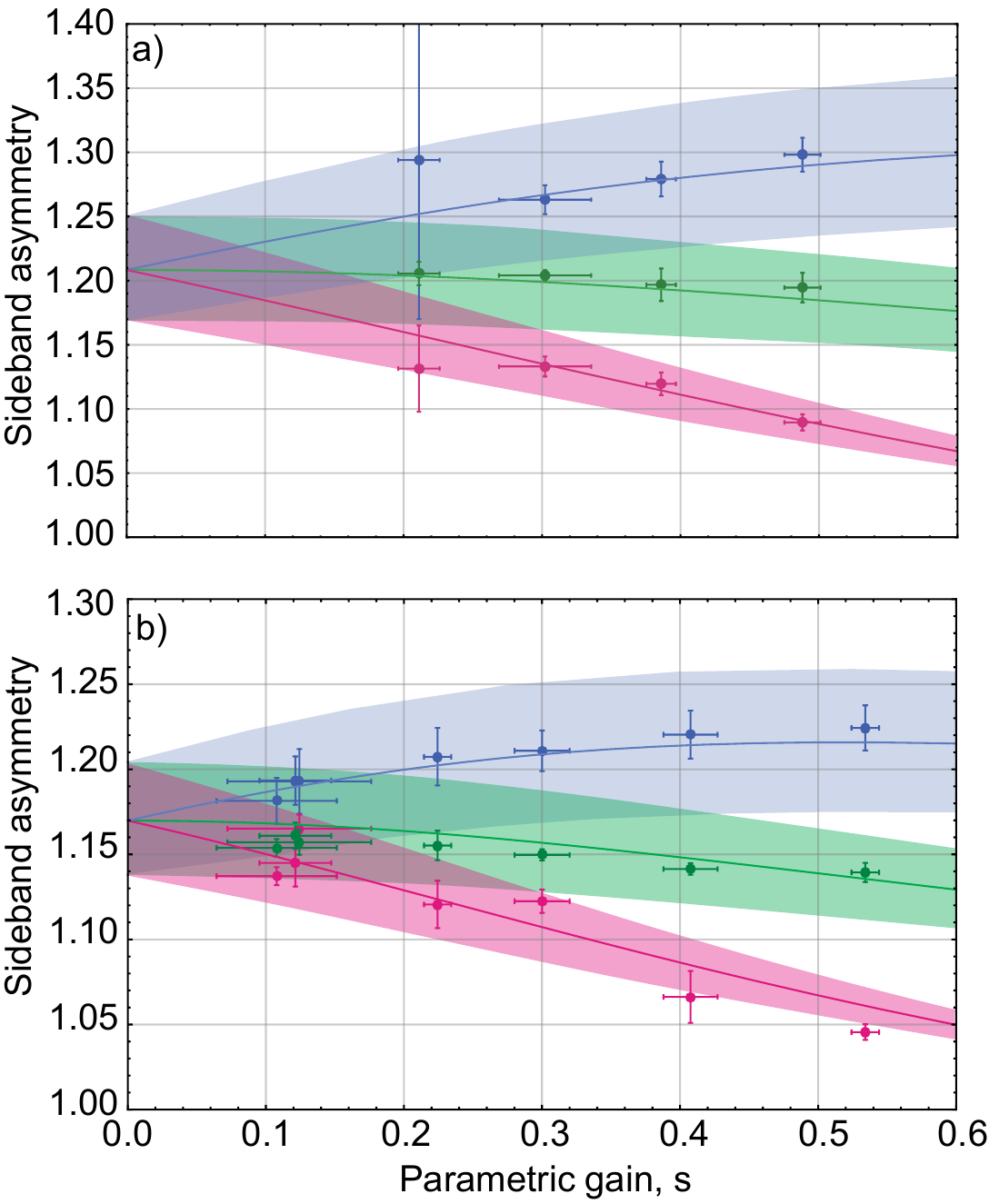}
\caption{Sideband asymmetry as a function of the parametric gain $s$ for constant detuning and for two
different occupation numbers (a) $\nm$ = $4.2$ and (b) $\nm$ = $5.8$. The green symbols refer to the sideband asymmetry $R_0$ without any resonant parametric drive, realized by detuning the modulation tone (see text). Blue (red) symbols indicate the ratio between the areas of the broad (narrow) Lorentzian contributions $\Rplus$ ($\Rminus$) in the two sidebands, observed with resonant parameteric modulation. The values of parametric gain $s$ are evaluated from the fitted widths $\Gplus = \Geff (1+s)$ and $\Gminus = \Geff (1-s)$. The error bars correspond to one standard deviation calculated from 5 consecutive independent measurements, each one lasting 100~s. The solid lines show the corresponding theoretical behavior, with shadowed areas given by the uncertainty in the system parameters ($5 \%$ in the cavity width and $0.5$ K in the temperature).}
\label{figure3}
\end{centering} 
\end{figure}

In Fig. \ref{figure3} we plot the sideband asymmetry at increasing strength of the parametric tone, keeping constant the total pump beam power, for two different values of the occupation number. When the parametric tone is not resonant (i.e., the parametric effect is off), the ratio $R_0$ (green symbols) remains almost constant for both occupation numbers, although we observe a clear decreasing trend as the parametric tone is increased. Such behavior is well reproduced by theoretical curves (see Eq. (\ref{eqGeff})) calculated by using independently measured parameters, and it is due to a change in the relative strength of the two pump tones, which results in a slightly reduced cooling power. We remind that the parametric tone is injected into the cavity through the pump beam and the parametric effect is turned off by shifting the driving frequency from $2\Omega_{m}$ to $2\Omega_{m}+\Dshift$ (see Sec. \ref{Sec:setup}). Although this procedure allows to reduce to a minimum the changes in the cooling efficiency (as explained in Sec. \ref{Sec:setup}), a residual effect is still present.

Fig. \ref{figure3} also shows the sidebands asymmetry for the Lorentzian components related to the broad quadrature  $R_{+}$ (blue symbols) and narrow quadrature $R_{-}$ (red symbols), with resonant parametric effect. The parametric gain $s$ used to calibrate the horizontal axis in the figure is deduced from the width of the Lorentzian curves, as above described. The corresponding theoretical ratios can be calculated from the theoretical spectra $S_{\bd \bd}$ and $S_{\bh \bh}$ (Eqs. \ref{eq_left}-\ref{eq_right}) and are given by Eqs. (\ref{ratios1}-\ref{ratios2}). Such theoretical curves are also plotted in Fig. \ref{figure3} without free fitting parameters, showing a good quantitative agreement with the experimental data.

\begin{figure}
\begin{centering}
\includegraphics[width=0.75\columnwidth]{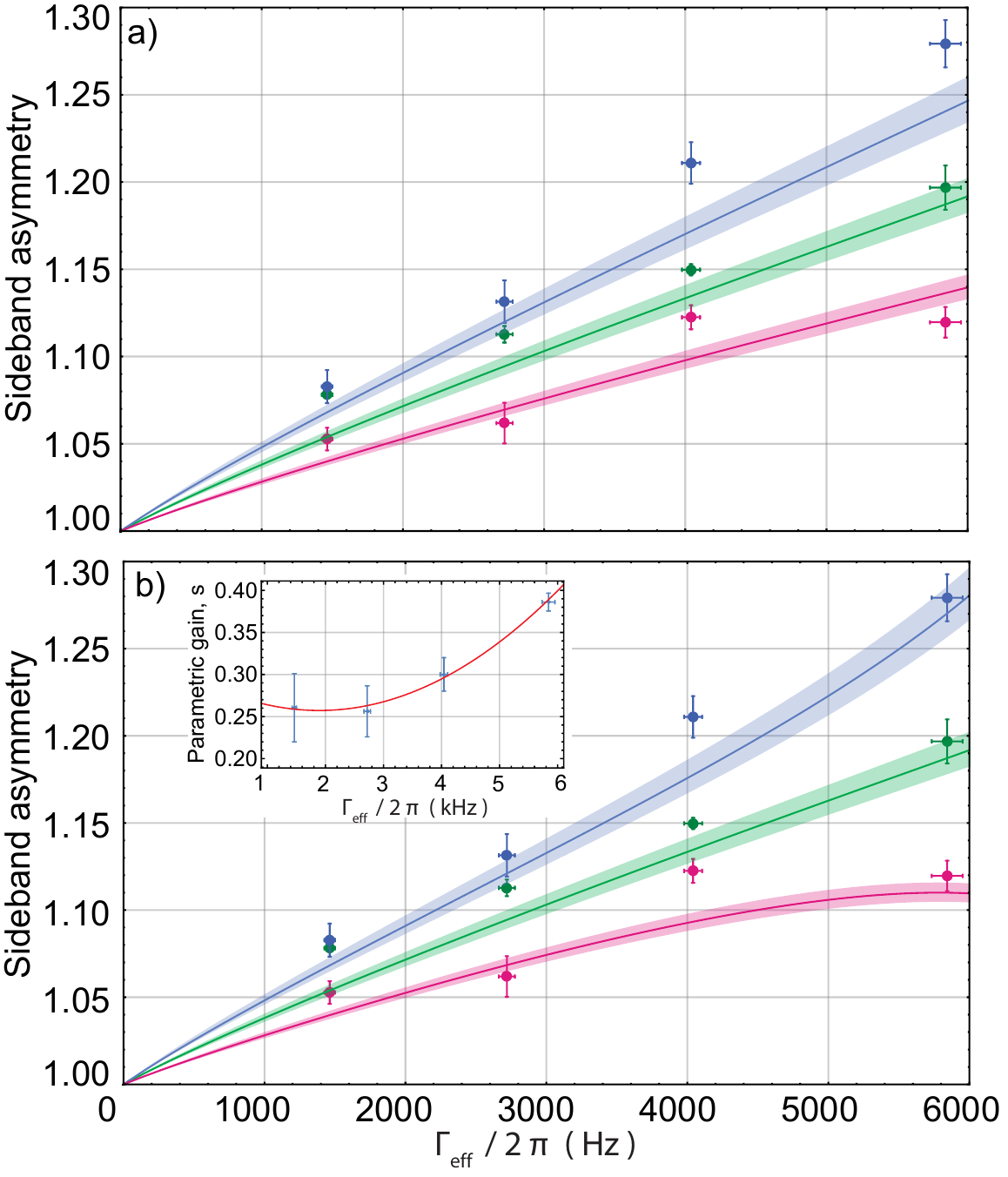}
\caption{(a) Sideband asymmetry as a function of $\Geff$ for fixed mean pump detuning $\Dpump$ and parametic gain $s$. Green symbols refer to the sideband asymmetry $R_0$ (without resonant parametric drive) while blue (red) symbols show ratios between the areas of the broad (narrow) Lorentzian contributions $\Rplus$ ($\Rminus$) in the presence of resonant parameteric modulation. The values of $\Geff$ in the abscissa are obtained from fits of the heterodyne spectra without resonant parametric drive. The solid lines show the corresponding theoretical behavior and shadowed areas represent the uncertainty in the system parameters (as in Fig. \ref{figure3}). The mean sideband asymmetry and their standard deviations are extracted from 5 consecutive independent measurements, each one lasting $100$~s. (b) The same experimental data are compared with modified theoretical curves taking into account the changes of $s$ with $\Geff$ (see inset).}
\label{figure4}
\end{centering} 
\end{figure}

We now analyze the sideband asymmetry as a function of the effective mechanical width $\Geff$, which is varied by increasing the pump power while keeping a constant mean detuning of the pump tones $\Dpump$, and parametric gain $s$. The experimental results and their relative theoretical curves are displayed in Fig. \ref{figure4}. 
In the absence of any parametric modulation, $R_0$ increases with $\Geff$ as expected. The mean phonon number in $R_0$ includes not only the cooling effect resulting from the optomechanical interaction, but also the additional occupation number due to the backaction from the probe and the pump beam.
When the parametric modulation is turned on, the equivalence between $\Rplus$ and $\Rminus$ is broken and the curves start to diverge, with a separation which increases with the pump power and thus with $\Geff$. 
While in the absence of parametric effect the agreement between experiment and theory is good, the data of $\Rplus$ and $\Rminus$ depart from the theoretical predictions, in particular at high pump powers.
We remark that the parametric gain can be written as $s=\Gpar/\Geff$ and thus is explicitly dependent on the mean pump detuning $\Dpump$ and on the ratio between the cooling tone and total pump power (see Eq. \ref{eq_Gpar}).
During the measurements, the parametric gain is maintained roughly constant by adjusting the strength of the parametric tone each time the pump power is varied.
Nevertheless, we observe a residual variation of $s$ as shown in the inset of Fig. \ref{figure4}b. We attribute such deviations to small changes of the locking point as the pump power increases, which induce changes in $\Dpump$ and hence of the parametric effect. In Fig. \ref{figure4}b we show the experimental data together with modified theoretical curves for $\Rplus$ and $\Rminus$, in which we phenomenologically include the dependence of $s$ on $\Geff$. These new theoretical lines show indeed a better agreement at high pump powers.

\begin{figure}
\begin{centering}
\includegraphics[width=0.8\columnwidth]{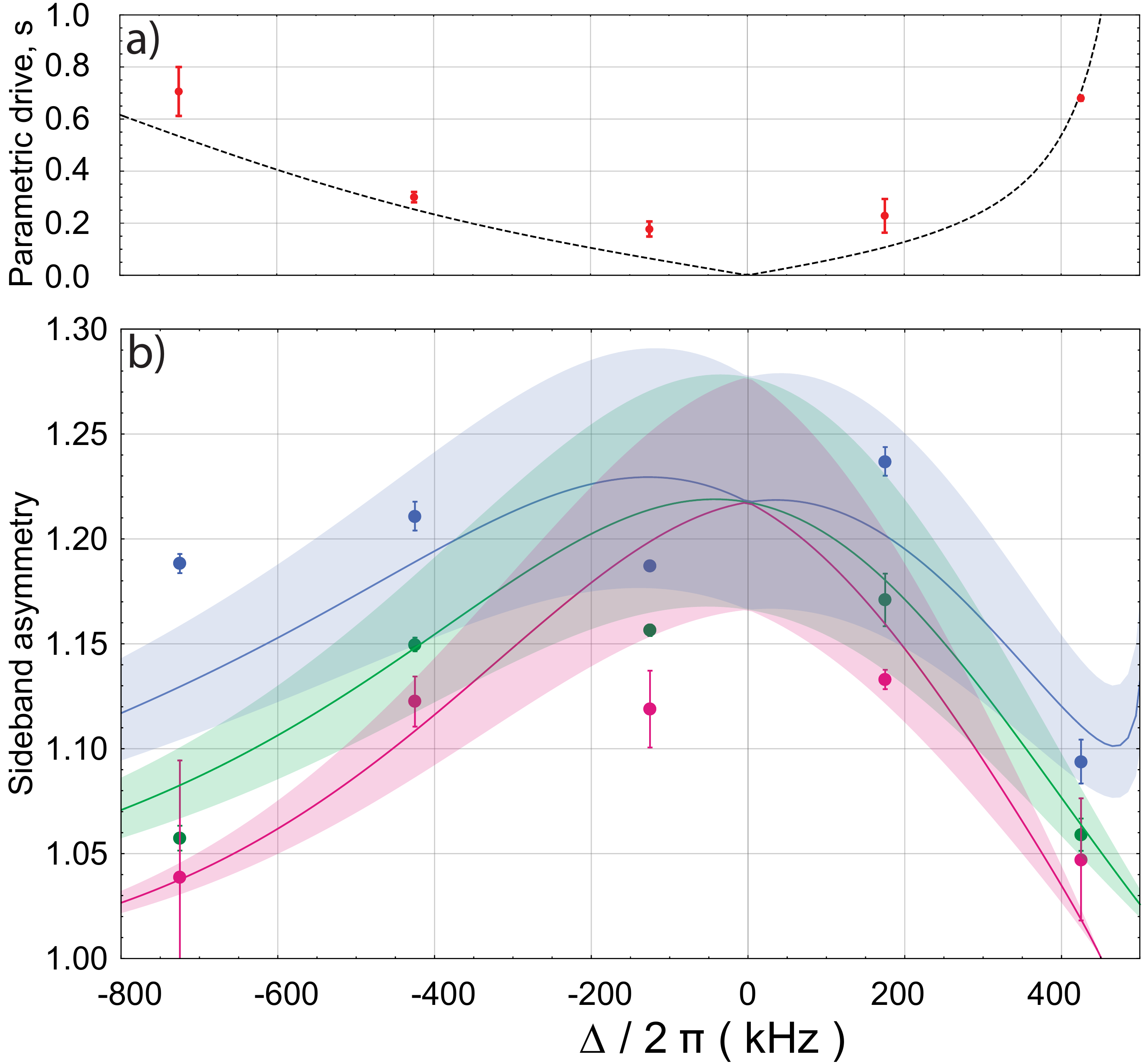}
\caption{(a) Parametric gain $s$ as a function of the mean detuning of the pump tones. The experimental values of the parametric gain (red symbols) are obtained from the fitted widths $\Gplus = \Geff (1+s)$ and $\Gminus = \Geff (1-s)$. The theoretical curve (dotted line) is $s=\Gpar/\Geff$ with $\Gpar$ and $\Geff$ given by Eqs. (\ref{eq_Gpar}) and (\ref{eqGeff}), respectively.
(b) Sideband asymmetries $R_0$ (green), $\Rplus$ (blue) and $\Rminus$ (red) as a function of the mean detuning of the pump tones $\Dpump$. The solid lines show the corresponding theoretical behavior and shadowed areas represent the uncertainty in the system parameters as in Fig. \ref{figure3}.}
\label{figure5}
\end{centering} 
\end{figure}

We finally study the dependence of the parametric gain and sideband asymmetry on $\Dpump$.
Its variation has the twofold effect of affecting the cooling efficiency and the parametric gain.
The former is evidenced by the variation of the sidebands asymmetry $R_{0}$ plotted in Fig. \ref{figure5}b (green points). The corresponding theoretical curve (green line) exhibits a maximum slightly below $\Dpump=0$. $\Dpump=0$ means that the cooling tone detuning is $\omega_{\mathrm{cool}}-\omega_{\mathrm{c}} = -\Omegam$, close to the value where optimal cooling is indeed expected.
%$\Delta_{pump}=(\Delta_{cool}+\Delta_{par})/2$
The parametric gain values are shown in Fig \ref{figure5}a together with the theoretical curve showing a minimum equal to zero at $\Dpump=0$. The parametric gain can indeed be written as $s=\Gpar/\Geff$ where $\Gpar \propto \Dpump$  as $\Dpump \rightarrow 0$.  Accordingly, as the null pump detuning is approached, the difference between the sideband ratios $\Rplus$ and $\Rminus$ decreases to disappear at $\Dpump=0$ (see Fig \ref{figure5}b).

\begin{figure}
\begin{centering}
\includegraphics[width=0.8\columnwidth]{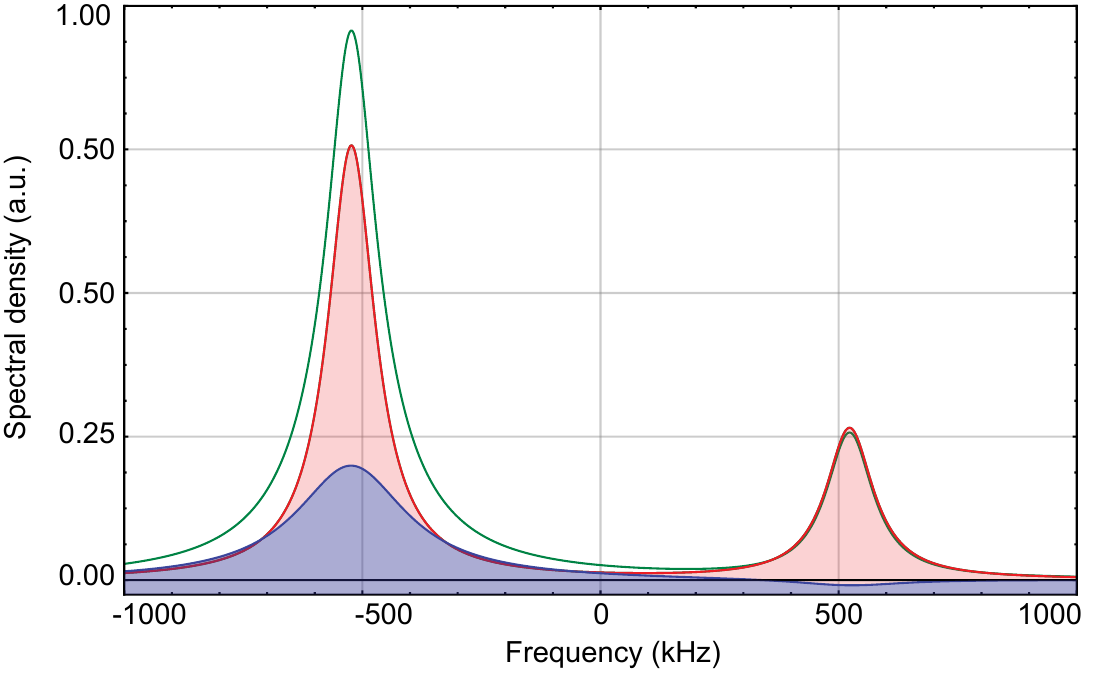}
\caption{Theoretical spectra of the Stokes and anti-Stokes sidebands as obtained from Eqs. (\ref{eq_left}-\ref{eq_right}) for $\nm = 0.12$ and $s=0.4$ (green solid line). Blue and red lines identify the broad and narrow Lorentzian components.}
\label{figure6}
\end{centering} 
\end{figure}

We have seen that sidebands spectra provide a powerful quantum indicator of a squeezed state: the 
narrow and broad Lorentzian components of each motional sideband gives a signature of the imbalance between the fluctuations in the two quadratures, while the sidebands asymmetry quantifies their non-classical nature.
Similarly to thermal states, even for the squeezed state the transition between classical and quantum behavior is smooth and some level of quantum squeezing is present even in macroscopic oscillators dominated by thermal noise. We remark indeed that the sideband asymmetry is in itself a fully quantum feature.
On the other hand, it is interesting to explore the extreme case in which the residual fluctuations in the squeezed quadrature are reduced below the zero-point level. This occurs for $(2\nm+1)/(1+s) < 1$, i.e., for $s > 2\nm$. Eq. (\ref{eq_right}) dictates that the broad Lorentzian contribution to the anti-Stokes sideband becomes negative, although this is over compensated by the narrow component, as the overall spectral density obviously remains positive at all frequencies. A spectrum with these characteristics is illustrated in Fig. \ref{figure6}. 
The negativity of the broad component in the anti-Stokes sideband provides a clear indication of a \emph{bona fide} quantum squeezing without the necessity of absolute calibrations. Under a continuous parametric drive the system is stable for $s < 1$, which represents the parametric instability threshold. The condition $s > 2 \nm$ would then require an initial occupation number $\nm < 0.5$, a level that has already been reached even in opto-mechanical setups based on SiN membranes  (see, e.g., Refs. \cite{pete2016,Galinskiy2020}). 
%On the other hand, such prerequisite could also be circumvented exploiting experimental techniques based on weak measurements and feedback, that have been demonstrated to allow stable operation even beyond the instability threshold \cite{Pontin_PRL2014,Sonar2018,Bowen2011,Bowen2013,Poot2014,Poot2015}.

\section{Conclusions}
\label{Sec:concl}

We have recently described a cavity opto-mechanics experiment where a macroscopic mechanical oscillator is parametrically driven by a suitable combination of optical fields \cite{ChowPRL2020}. We have shown that the generated mechanical squeezed state exhibits a quantum dynamics that is evidenced by the shape of the motional sidebands. In the present work we describe the theoretical model behind this phenomenon, and present a detailed characterization of the experimental achievements, that is in good agreement with the model. We suggest that the analysis of the motional sidebands can provide a clear signature of the noise reduction below the zero-point fluctuations that occurs in one quadrature, without requiring any absolute calibration of the displacement spectra, nor even a direct measurement of the quadrature fluctuations. 

Our results widen the range of macroscopic non-classical states that can be explored in opto-mechanical experiments. For instance, interesting developments can involve non-stationary squeezed states and multimode squeezing \cite{MahboobPRL2014,Patil2014,Pontin2016}.

\section*{Acknowledgments}

Research performed within the Project QuaSeRT funded by the QuantERA ERA-NET
Cofund in Quantum Technologies implemented within the European Union' s
Horizon 2020 Programme. The research has been partially supported by INFN
(HUMOR project).

\end{document}